\documentclass[a4paper,11pt]{article}
\pdfoutput=1 

\usepackage{jcappub} 
\usepackage{amsmath}
\usepackage{natbib}
\usepackage[T1]{fontenc} 

\usepackage{graphicx}
\graphicspath{ {Images/} }

\title{\boldmath  Energy injection in pre-recombination era  and  EDGES detection}

\author[{a,1}]{Nirmalya Brahma,\note{Corresponding author.}}
\author[b]{Shiv Sethi,}
\author[a]{Shivnag Sista}

\affiliation[a]{Indian Institute of Science, \\
C.V.Raman Avenue, Bangalore, Karnataka 560012, India}
\affiliation[b]{Raman Research Institute, \\ C.V.Raman Avenue, Bangalore, Karnataka 560080, India }

\emailAdd{nirmalyab@iisc.ac.in}
\emailAdd{sethi@rri.res.in}
\emailAdd{shivnag@iisc.ac.in}

\abstract{We study the possibility of explaining the recent EDGES detection by an energy injection in the pre-recombination era. Our aim is to show that the residue of this energy injection could give the resultant increase in the energy density at frequencies $x_e \equiv h\nu/(kT) \simeq 10^{-3}$, which is needed to explain the EDGES result. We consider two models of energy injection: Gaussian profile with a fixed redshift of injection and radiative decay of a  non-relativistic particle.   We show that the energy injection should occur in the redshift range $z\lesssim 4 \times 10^3$ to prevent free-free processes  from thermalizing the injected energy. The injected energy should be nearly 200--1000 times the CMB intensity  at $x_e \simeq 10^{-3}$ to obtain  the requisite residue. A large fraction of the injected energy gets thermalized and therefore distorts the CMB spectrum. We compute CMB spectral distortion for both the models and show that the fractional  change in CMB energy density,  $\Delta \rho_{\rm \scriptscriptstyle CMB}/\rho_{\rm \scriptscriptstyle CMB} \simeq 10^{-6}$, which might be detectable with the  proposed experiment PIXIE. We also outline  the implication of our proposed scenario for CMB anisotropies.}

\begin{document}
\maketitle
\flushbottom

\section{Introduction}

In recent years, major advances have been made in our understanding
of the formation of structures in the universe via precise  cosmic microwave background (CMB) anisotropy experiments \cite{Planck2018,Ade:2015xua,Hinshaw:2012aka} and  large scale structure surveys \cite{Beutler:2016ixs}. Another major cornerstone of modern cosmology is the near-blackbody spectrum of CMB. The 
COBE-FIRAS experiment gave the current upper bounds on the CMB  spectral distortion parameters: $|\mu| \lesssim 9 \times 10^{-5}$ and $|y| \lesssim 1.5 \times 10^{-5}$ \cite{Fixsen:1996nj}. Such stringent upper limits on the deviation of CMB spectrum from a blackbody put strong constraints  on processes that inject energy
into the plasma in the universe after $z \simeq 10^6$ (for details e.g.
\cite{1977NCimR...7..277D,1991ApJ...371...14D,1991A&A...246...49B,Hu:1994bz,Khatri:2011aj,Chluba:2011hw,Chluba:2012gq,Khatri:2012tw,2012JCAP...06..038K,Dent:2012ne,Chluba:2012we,Chluba:2013wsa,2014PTEP.2014fB107T,Hill:2015tqa,chluba2019new,kogut2019cmb,Chluba:2020oip}). The proposed  experiment 
PIXIE \cite{Kogut:2011xw}  will improve FIRAS limits  by
many orders of magnitude:  $y \simeq  10^{-9}$ and $\mu \simeq  10^{-8}$. 

CMB physics  has also enhanced our understanding of the dark age of the universe which is  predicted to have ended  around redshift $z \simeq 35$ with the formation of first large-scale structures (epoch of cosmic dawn). This era was followed by the epoch of reionization (EoR) during which the UV photons from these  collapsed structures emitted radiation  which  heated and ionized their surrounding medium  until  $z \simeq 8$ (\cite{Barkana:2000fd,2010ARA&A..48..127M,21cm_21cen,2014PTEP.2014fB112N}).  The CMB temperature and polarization anisotropy detection by WMAP and Planck determine  the redshift of reionization,  $z_{\rm reion} = 7.75 \pm 0.73$ (\cite{Planck2013,Hinshaw:2012aka,Planck2015,Fan:2000gq,2001AJ....122.2850B,Planck2018}).

The cleanest   probe of  the physics of EoR is through the detection of redshifted hyperfine 21~cm line of neutral hydrogen (HI). This signal carries crucial information about the first sources of radiation in the universe and their spectrum in three frequency bands: ultraviolet (UV) radiation (ionizes the surrounding medium), Lyman-$\alpha$ radiation (determines the relative population of   neutral hydrogen atoms in hyperfine states), and x-ray photons (heat and partially ionize the medium). In addition, the sources that  emitted soft radio photons  would also affect the observable HI signal  (e.g. \cite{2018arXiv180301815E,feng18}).

The epochs of cosmic dawn and  EoR have  been  studied in detail  using numerical, semi-analytic   and,  analytic methods (e.g. \cite{2007MNRAS.376.1680P,21CMFAST,2012Natur.487...70V,2013MNRAS.435.3001T,2013MNRAS.431..621M,2014MNRAS.443..678P,LateHeat2,2015MNRAS.447.1806G,2017MNRAS.464.3498F,RS18,2019ApJ...876...56R}). Theoretical estimates based on standard thermal and ionization history  suggest  the global signal is observable in both absorption and emission with  strength $-200\hbox{--}20 \, \rm mK$ in a frequency range  $50\hbox{--}150 \, \rm MHz$,  corresponding  to a redshift range $25 > z > 8$ (e.g. \cite{1997ApJ...475..429M,2000NuPhS..80C0509T,2004ApJ...608..611G,Sethi05}). The fluctuating component of the signal is expected to be an order of magnitude smaller on scales in the range $3\hbox{--}100 \, \rm Mpc$, which implies  angular scales  $\simeq 1\hbox{--}30$~arc-minutes (e.g. \cite{ZFH04,FZH04a,FZH04b,2007MNRAS.376.1680P}; for comprehensive reviews see e.g. \cite{21cm_21cen,2014PTEP.2014fB112N,2010ARA&A..48..127M}). 

The recent detection of a broad global absorption trough of strength $500 \, \rm mK$ by the  EDGES group (\cite{EDGES2018}) at $\nu \simeq 80 \pm 10 \, \rm MHz$ is the only positive detection of HI signal at high redshifts. This
detection is in disagreement with the standard scenario and 
might be pointing to the presence of unknown physics in pre- or
post-recombination universe  (e.g. milli-charged dark matter or  strong radio background at $z\simeq 20$ \cite{Barkana2018,2018arXiv180405318L,2018PhLB..785..159F,2018Natur.557..684M,feng18,2018arXiv180301815E}). Such a  feature  could also  potentially be explained by absorption in the Galactic  interstellar medium \cite{draine-miralda}.

In this paper, we attempt to explain the EDGES results by an  injection of energy
in the Rayleigh-Jeans part of the CMB spectrum in the   pre-recombination era (for a detailed discussion see \cite{chluba2015green}). If the injected photons equilibrate before the recombination occurs,  there would be no trace of the injection event except global distortion of CMB spectrum. However, depending on the amount of injected energy and the time scale of  equilibration, there could be residue excess which might leave observable signatures  on the CMB spectrum close to energies of injection. This excess 
 can  explain the depth of EDGES' absorption feature. The possibility of such a signature depends on the competition between the rates of interaction between the photon and matter field, the expansion rate,  the frequency range,  and  redshifts at which the photons are injected. In this paper, we  address  this question  by evolving the CMB spectrum  after such an energy injection event, using both analytical and numerical methods. 

One distinct advantage of studying an energy injection  during  this era is that both CMB  anisotropies and spectral distortion strongly constrain the scope of this injection. Therefore, a generic episode of such
energy injection would, in addition to explaining the EDGES detection,  likely leave detectable signatures on CMB. 

In the next section, we review the HI signal from the EoR  and cosmic dawn,
the EDGES results, and its proposed  explanation. In section~\ref{sec:cmbspec} we discuss the physical processes involved in determining the CMB spectrum;
 the relevant  time scales are given in the Appendix. In section~\ref{sec:anaappro}, we discuss analytic
approaches  to solving the coupled evolution of the  photon occupation number (Kompaneets equation)  and electron
temperature. In section~\ref{sec:numsol}, the numerical solutions are presented
along with the particle decay model. In the final section, we summarize our findings and outline future prospects. 
Throughout this paper, we assume the spatially-flat $\Lambda$CDM model with the following parameters: $\Omega_m = 0.310$, $\Omega_B = 0.049$, and  $h = 0.677$ (\cite{Planck2018}).

\section{HI signal, EDGES detection, and possible explanations} \label{sec:edgesre}

In the  atom rest frame, hyperfine splitting of the ground state of neutral hydrogen (HI)
causes an energy difference that corresponds to a wavelength $\lambda = 21.1 \, \rm cm$. The excitation temperature of this line, $T_S$, is determined  by three processes in the early universe: emission and 
absorption of  CMB radiation  which is a blackbody of temperature $T_{\rm CMB}$, 
collisions with atoms, and the mixing of the two levels caused 
by  Lyman-$\alpha$ photons (Wouthuysen-Field effect \cite{wouthuysen1952excitation,field1959spin,Field1958}).  $T_S$ can be expressed in terms of 
the colour temperature of Lyman-$\alpha$ photons, $T_{\alpha}$,   gas kinetic temperature $T_K$, and $T_{\rm CMB}$ (\cite{Field1958,field1959spin,21cm_21cen}):
	\begin{equation}
		T_S=\frac{T_{\rm CMB}+y_{\alpha}T_{\alpha}+y_c T_K}{1+y_{\alpha}+y_c}
                \label{eq:tsbas}
	\end{equation}
Here $y_c \propto n_{\rm H}$ and $y_\alpha \propto n_\alpha$ ($n_{\rm H}$ and $n_\alpha$
are the number densities of neutral hydrogen atoms and Lyman-$\alpha$ photons,
respectively) determine the efficiency of  collisions and Lyman-$\alpha$ photons, respectively. In the early universe, $1000 < z < 100$, $T_S$ relaxes to $T_{\rm CMB}$. In the redshift range $100 < z < 30$, collisions determine the spin temperature and $T_S$ relaxes to the kinetic 
temperature $T_K$ of the matter. As the epoch of reionization commences, the production of Lyman-$\alpha$ photons couples the spin temperature to the colour 
temperature of Lyman-$\alpha$, $T_\alpha$. 
It can be shown that multiple scattering of Lyman-$\alpha$ photons with HI  causes $T_\alpha$ to relax to the kinetic temperature (e.g. \cite{2004ApJ...602....1C,1959ApJ...129..551F,1994ApJ...427..603R}). 
Therefore if $y_{\text{tot}}=y_c+y_{\alpha} \gtrsim T_{\rm CMB}/T_K $, then $T_S \simeq T_K$. Otherwise, it relaxes to $T_{\rm CMB}$. 

The HI  emits or absorbs 21-cm radiation from CMB depending on whether its excitation temperature  $T_S$ is greater than or less than $T_{\rm CMB}$. This temperature difference is observable and, averaged over the sky,  can be expressed as (e.g. \cite{21cm_21cen,1997ApJ...475..429M,1999A&A...345..380S,2004ApJ...608..611G,Sethi05}):
	\begin{align}
	  \Delta T_b  &\simeq \frac{\tau}{1+z}(T_S-T_{\text{CMB}}) \nonumber \\
				& \simeq 26.25 f_{\rm HI}\;\left(1-\frac{T_{\text{CMB}}}{T_S}\right) \left(\frac{1+z}{10}\frac{0.14}{\Omega_m h^2}\right)^{\frac{1}{2}} \left(\frac{\Omega_b h^2}{0.022}\right) \text{mK} \label{overallnorm}
	\end{align}
Here $f_{\rm HI}$ is the fraction of gas in the neutral phase.

Recent EDGES observation (\cite{EDGES2018}) reported  a sky-averaged absorption feature of strength $\Delta T \simeq -500 \,\rm mK$ in the frequency range $70$--$90$~MHz, corresponding  to a redshift range $15$--$19$ for the redshifted HI line. It can be shown that, for standard recombination and thermal history, the minimum temperature of the gas at $z\simeq 19$ is $T_K \simeq 6 \, \rm K$. It follows from Eqs.~(\ref{eq:tsbas}) and~(\ref{overallnorm}) that the absorption trough should not have been deeper than $-180 \, \rm mK$. 

One possible explanation of the EDGES result is  additional  radio background in the redshift range  $15 < z< 19$ whose brightness temperature $T_{\rm radio}$ is higher than the CMB temperature; in this case we can replace $T_{\rm CMB}$ with the $T_{\rm CMB} +T_{\rm radio}$ in Eq.~(\ref{eq:tsbas}) in the relevant  redshift range (\cite{feng18}, \cite{2018arXiv180301815E}, \cite{sharma18}). With this replacement and suitable choice of $T_{\rm radio}$ the EDGES result can be explained.


Another plausible explanation invokes the additional cooling of baryons owing to interaction between dark matter and baryons \cite{Barkana2018}.  In this case, we can explain the EDGES detection using  Eqs.~(\ref{eq:tsbas}) and~(\ref{overallnorm}) if the  dark matter-baryon coupling can cool the baryons such that the matter temperature  $T_K  \lesssim 2.5 \, \rm K$ in the redshift range of interest  \footnote{EDGES detection implies a sharp trough in the signal at $z\simeq 19$ and a sharp rise at $z\simeq 15$. As the noise level for the detection is $\simeq 20 \, \rm mK$ \cite{EDGES2018}, the trough  at higher redshift can arise from complete Lyman-$\alpha$ coupling being established close to $z\simeq 19$ with the  rapid heating being responsible for the sharp rise at smaller redshift (e.g. \cite{2019ApJ...876...56R})}.  Other possible explanations of this result include a possible systematic error \citep{hills18} and  absorption  from spinning dust grains in the  Galactic ISM  \citep{draine-miralda}.

In summary, the EDGES result can be explained by raising $T_{\rm CMB}$ or reducing $T_S$ by a factor of roughly 2.5 in Eq.~(\ref{overallnorm}). Given the
uncertainty in other physical processes such as Lyman-$\alpha$ coupling and
x-ray heating, this factor provides a lower limit on the required enhancement/reduction in the relevant frequency range (e.g. \cite{2019ApJ...876...56R}). 

In this paper, we invoke pre-recombination physics to explain the EDGES result.
In particular, we note that the excess radio background needed to explain
the EGDES result could be a relic of the pre-recombination era.

We consider energy  injection of suitable amplitude and   frequency range
to explain the EGDES
result. In the pre-recombination era, these photons are subject to multiple
physical processes which can upscatter (inverse Compton scattering),  downscatter (Compton scattering), or absorb (free-free or double Compton processes) these photons. We study the impact of all these processes  to
discern the range of redshifts at which these photons can be injected. We also
study the impact of these photons on other observables such as spectral
distortion of CMB and CMB anisotropies. In the next section, we review these
physical processes.

\section{CMB spectrum: Physical processes} \label{sec:cmbspec}

We assume that  the energy is  injected in a narrow range of frequencies
into CMB at an early epoch of the universe. Our aim is to 
study how this distorted spectrum  evolves with time. The redshift
range of injection  for our study is  $10^3 < z < 10^{5}$.
The reason for this choice has been elaborated upon in  later sections.
We need to understand   the relative importance of  various radiative
processes that are responsible for the time evolution of the spectrum.

The photon occupation number $\eta(\nu,t)$ is a function of
the frequency (or equivalently, the energy) and  time. The following  physical processes  impact the evolution of the photon occupation number:
\begin{itemize}
\item Compton and inverse Compton  scattering
\item Double Compton emission and absorption
\item Free-free emission and absorption
\end{itemize}
Compton scattering (and its inverse process) only change the energy
of photons without affecting the photon number while the other two
processes can absorb and create photons. 

The rate of change of the photon occupation number due to Compton and inverse Compton scattering off non-relativistic electrons  is given by Kompaneets equation (\cite{1957JETP....4..730K, sunyaev1970interaction, zel1972effect, illarionov1974comptonization, pozdnyakov1983astrophys}):
\begin{equation}
  \left(\frac{\partial\eta}{\partial t}\right)_{C}=\frac{a_{c}}{x_{e}^{2}}\frac{\partial}{\partial x_{e}}\left(x_{e}^{4}\left[\frac{\partial\eta}{\partial x_{e}}+\eta+\eta^{2}\right]\right)
  \label{eq:comp_invcomp}
\end{equation}
Here $x_{e}=h\nu/kT_{e}$ is a dimensionless parameter, independent of redshift for equilibrium $\eta$,  expressed in terms of  the physical (not comoving)  frequency $\nu$ and the electron temperature $T_{e}$ \footnote{The electron temperature is the same as matter temperature throughout the pre-recombination era as the physical processes that equilibrate energy between electrons and baryons act on time scales far shorter than the expansion time scale.}; $a_{c}=n_{e}\sigma_{T}c\left(kT_{e}/(m_{e}c^2)\right)$, where
$n_{e}$ is the number density of electrons, $\sigma_{T}$ is the Thompson
cross-section and $m_{e}$ is the mass of an electron. 
The $\partial\eta/\partial x_{e}$ term in the  equation corresponds
to the contribution for inverse Compton scattering, which results
in photons gaining energy from heated electrons. The $\eta$ and $\eta^{2}$ terms on the right-hand
side of the  equation correspond to spontaneous and stimulated
scattering, respectively, both of which contribute to the cooling of
photons and are collectively referred to as Compton scattering.

The time evolution of the photon occupation number from  double Compton processes---emission, absorption, and stimulated emission---is given by:
\begin{equation}
\left(\frac{\partial\eta}{\partial t}\right)_{DC}=\frac{C(t)}{x_{e}^{3}}\left[1-\eta\left(\exp(x_{e})-1\right)\right]\label{eq:double_comp}
\end{equation}
 Here $C(t)=(4\alpha/(3\pi))(kT_{e}/(mc^{2}))^{2}(I(t)/t_{c})$, $\alpha=e^{2}/\hbar c$ is the fine structure constant and $t_{c}=(n_{e}\sigma_{\rm T}c)^{-1}$
is the photon-electron collision time.
\begin{equation}
I(t)=\intop_{0}^{\infty}x_{e}^{4}(1+\eta)dx_{e}
\end{equation}
We follow the prescription of \cite{1982A&A...107...39D} in estimating $I(t)$, which gives us $I(t) \simeq 4\pi^4/15$ (for more discussion on this
approximation see \cite{2007A&A...468..785C}). 

Similarly, the contribution of all the relevant free-free  processes to the evolution of photon occupation number is given by (\cite{sunyaev1970interaction, zel1972effect, illarionov1974comptonization}):
\begin{equation}
  \left(\frac{\partial\eta}{\partial t}\right)_{B}=K_{0}\frac{g(x_{e})\exp(-x_{e})}{x_{e}^{3}}\left(1+\eta(x_{e})(1-\exp(x_{e}))\right)
  \label{eq:freefree}
\end{equation}
where $g(x_{e})$ is the Gaunt factor.  For  $x_e \ll 1$, $g(x_{e}) \simeq \sqrt{3}/\pi\ln(2.25/x_e)$ (e.g. \cite{1977NCimR...7..277D}; for a detailed discussion on the Gaunt factor see e.g. \cite{1979rpa..book.....R}).
The expression for $K_0$  is given in Eq.~(\ref{eq:freefree_eabs}).

Combining the contributions from Compton, double Compton and free-free
processes, we obtain the generalized Kompaneets equation:
\begin{multline}
  \left(\frac{\partial\eta}{\partial t}\right)=\frac{a_{c}}{x_{e}^{2}}\frac{\partial}{\partial x_{e}}\left(x_{e}^{4}\left[\frac{\partial\eta}{\partial x_{e}}+\eta+\eta^{2}\right]\right)+\frac{C(t)}{x_{e}^{3}}\left[1-\eta\left(\exp(x_{e})-1\right)\right]\\+K_{0}\frac{g(x_{e})\exp(-x_{e})}{x_{e}^{3}}\left(1+\eta(1-\exp(x_{e}))\right)
  \label{eq:kompanee}
\end{multline}
This equation is solved along with the evolution of the  temperature, $T_e$, which is  the temperature shared by all the baryons  in the pre-recombination era (e.g. \cite{1977NCimR...7..277D}):
\begin{equation}
  {d T_e \over dt} = - {2\dot a \over  a}T_e + {(T_{\rm eq} - T_e) \over t_{e\gamma}} + {q_{\rm ff} + q_{\rm dc} \over 3 n_e k}
  \label{eq:elec_temp}
\end{equation}
Here $t_{e\gamma} = (3m_ec/4\sigma_T \rho_\gamma)$ and $T_{\rm eq}$, the  effective  temperature for an arbitrary photon occupation number $\eta$, is (\cite{zel1970stationary, levich1970heating}):
\begin{equation}
  T_{\rm eq} = {h \int_0^{\infty} d\nu  \nu^4 \eta (\eta+1) \over 4 k \int_0^{\infty} d\nu  \nu^3 \eta}
  \label{eq:teq}
\end{equation}
$q_{\rm ff}$ and $q_{\rm dc}$ give  the heating/cooling rates of the medium owing to
free-free and double Compton processes:
\begin{eqnarray}
  q_{\rm ff} & = & {8\pi (kT_e)^3 \over c^3 h^2} K_0 \int_0^\infty g(x_e) \exp(-x_e)\left[1+\eta(1-\exp(x_e))\right] d\nu \\
  q_{\rm dc} & = & {8\pi (kT_e)^3 \over c^3 h^2} C(t) \int_0^\infty \left[1-\eta(\exp(x_e)-1)\right] d\nu
  \label{eq:temp_ffdc}
  \end{eqnarray}

Before the energy is injected, the baryons and photons share a common
temperature with equilibrium photon occupation number given by the Planckian:
\begin{equation}
  \eta(x_e) = {1\over \exp(x_e) -1}
  \label{eq:eq_ph}
\end{equation}
It can readily be checked that  for this form of equilibrium photon occupation  
number the right hand sides of Eqs.~(\ref{eq:kompanee}) and~(\ref{eq:elec_temp}) vanish \footnote{The RHS vanishes for only this function. However, if only Compton
  and inverse Compton scatterings are considered, another equilibrium solution
  is Bose-Einstein distribution function with non-zero chemical potential}
and Eq.~(\ref{eq:teq}) yields $T_{\rm eq} = T_e$. In an expanding universe,
Eq.~(\ref{eq:eq_ph}) is left unchanged as $x_e$ is an invariant at early times.
After the injection of energy, $x_e$ is not an invariant and we can choose
another variable which remain invariant during the expansion of the universe (see below).

The time scales of the physical processes discussed above  are given in the Appendix (section~\ref{sec:tscale}).

\section{Analytical Approach and Approximations} \label{sec:anaappro}
Our aim is to solve Eqs.~(\ref{eq:kompanee}) and~(\ref{eq:elec_temp}) simultaneously after the injection of photons in a narrow frequency range at $x_e \ll 1$. We present
numerical solutions in the next section. Many of the results we find 
in this and the next section can also be obtained by an alternative approach
developed by Chluba \cite{chluba2015green}. In this section, we seek an
analytic approach based on the physical setting we propose in this
paper. Its salient points can be summarized as:
\begin{itemize}
\item[1.] The energy is injected in a frequency range for which $x_e \ll 1$. To explain the EDGES detection  we require a residue of the energy  injection in the range:  $x_e \simeq 1.2\hbox{--}1.6\times 10^{-3}$.
\item[2.] The amount of injected energy is much smaller than the energy
  density of the CMB, $\delta \rho_\gamma \ll \rho_{\rm \scriptscriptstyle CMB}$. If the energy
  is injected  for a range of frequencies between $x_f\equiv h\nu_f/(kT)$ and $x_i$ such that $x_f \ll 1$, then
  \begin{equation}
    \delta\rho_\gamma \simeq  {8\pi (kT)^4 \over 3 c^3 h^3}\left (x_f^3-x_i^3 \right ) \left ({T_B \over T}\right)
    \label{eq:eneinj}
  \end{equation}
  Here $T_B$ is the brightness temperature of the injected photons and
  $T$ refers to the equilibrium  matter/radiation temperature. The CMB energy density,   $\rho_{\rm \scriptscriptstyle CMB} = a T^4$, where $a = 8\pi^5 k^4/(15 h^3 c^3)$  is the radiation constant. For  $x_f = 2\times 10^{-3}$,   $\delta\rho_\gamma \simeq (T_B/T) x_f^3 \rho_{\rm \scriptscriptstyle CMB} \simeq 10^{-9} \rho_{\rm \scriptscriptstyle CMB}$ for  $T_B=T$. 
\item[3.] At these frequencies, both double Compton and free-free processes
  play an important role, as their time   scales are $ \propto x_e^2$. Double Compton
  processes  are more important as compared to free-free emission/absorption at higher redshifts as  they  scale as  $(1+z)^5$ while free-free processes scale as $(1+z)^{5/2}$. For $z < 10^{4}$, the free-free absorption is the dominant
  process (Section~\ref{sec:tscale}). 
\end{itemize}
As the efficiency  of double Compton and free-free processes is a sharp
function of frequency, it partly allows us to isolate the impact of these processes from  Compton  scattering.

We first consider the case for which double Compton and free-free processes are not efficient. It is further assumed that  the brightness  temperature of photons, $T_{\rm B} \gg T_e$, or
the energy density of injected photons far exceeds the energy density
of the equilibrium Planckian in the relevant frequency range. For $x_e \ll 1$, the equilibrium photon occupation number, 
$\eta \simeq 1/x_e \gg 1$. With the injection of additional photons, the photon occupation number becomes $\eta \simeq  1/x_B$, with $x_B = h\nu/(kT_B)$. As $x_B \ll x_e$, the $\eta^2$ term (the term corresponding to stimulated emission) in Eq.~(\ref{eq:comp_invcomp}) dominates the other two terms and  Eq.~(\ref{eq:comp_invcomp}) reduces to (\cite{zel1969bose, illarionov1972compton, syunyaev1971induced,1971ApJ...164..457A}): 
\begin{equation}
  \left(\frac{\partial\eta}{\partial y}\right)_{C}=\frac{1}{x_{e}^{2}}\frac{\partial}{\partial x_{e}}\left(x_{e}^{4}\eta^{2}\right)
  \label{eq:comp_sh}
\end{equation}
After  substitution: $f(x_e,y)=x_{e}\eta^{2}$, this equation reduces to
Burgers equation, which corresponds to the formation of a one-dimensional shock,  with solution (see also \cite{1977NCimR...7..277D}): 
\begin{equation}
f=g(x_{e}+2yf)\label{eq:20}
\end{equation}
where g(...) is any well-behaved function which provides the initial profile
of the injected photons.  The behaviour  of Eq.~(\ref{eq:20}) can be gauged from the well-known solution to Burgers equation. It corresponds to the excess photon profile moving towards lower frequency with a speed $2fa_c$. The speed  is higher for larger value of  $f$, just like the passage to a  shock solution. This results in a steepening of the initial profile which cause the 
 $d\eta/dx_e$ term  to become important in Eq.~(\ref{eq:comp_invcomp}).

In other words,  if the  injected energy dominates the background intensity  in a narrow range of frequencies such that $\eta \gg 1/x_e$ and its profile is smooth, the dominant
process that determines the initial evolution of the profile  is the Compton scattering, which arises from the recoil of electron in its rest frame. The Compton scattering   causes the photons to lose energy to electrons  at a   rate  $f a_c$, which depends on both the height of the profile and the Compton energy exchange time scale. However, the validity of Eq.~(\ref{eq:comp_sh}) breaks down after the initial phase 
because its solution (Eq.~(\ref{eq:20})) results in the steepening of the profile. As the profile steepens, the $d\eta/dx_e$ term  in  Eq.~(\ref{eq:comp_invcomp}) cannot be neglected. As noted above, this term corresponds to energy
exchange owing to the motion of electrons in the lab frame (inverse Compton scattering) and its net impact is to heat the photons.  When the inverse Compton scattering becomes important,  the photons start gaining energy. (This situation
is analogous to the solution of Burgers equation for the  one-dimensional shock solution. When velocity gradients become large close to the shock front, the viscosity cannot be neglected, resulting in the dissipation of energy and smoothening of the shock profile.) 

The effect of these two competing processes is to  shift a majority of excess photons to smaller frequencies during the initial
phase, followed by the  smoothening of  the   profile which 
upscatters  a fraction of  the photons to higher frequencies\footnote{It is easier to see how the two terms in Eq.~(\ref{eq:comp_invcomp}) compete if $x_e$ is explicitly written in terms of $\nu$ and electron temperature $T_e$. In equilibrium $T_e = T_{\rm CMB}$ these terms cancel each other. If $T_e > T_{\rm CMB}$, the inverse Compton scattering term dominates. In our case, $T_e \simeq T_{\rm CMB}$ even after the energy injection as these temperatures are determined by ratio of injected and equilibrium energy density which is very small. Therefore, in our case, the inverse Compton scattering becomes important  owing to  steepening of the $\eta$ profile. This also allows us to see how equilibrium is attained when only Compton/inverse Compton processes operate. }. In the inverse Compton scattering, the photons gain energy as $\propto \exp(y)$.  As $y \simeq 1$
at $z\simeq 2 \times 10^5$ and it scales as $(1+z)^2$ (in radiation-dominated era) this gain could be  negligible if the energy
is injected at a later redshift.  If the injection of the energy is at redshifts such that $y <1$, the inverse Compton  scattering cannot populate photons at  frequencies  much larger than the frequency range of injected photons.   This idealized case gives us important insights into the numerical solutions of Kompaneets equation and we will compare our numerical solutions with Eq.~(\ref{eq:20}) in the next section.

The absorption and emission of photons owing to double Compton
and free-free processes (Eqs.~(\ref{eq:double_comp}) and~(\ref{eq:freefree})) also have significant impact on the evolution of the photon occupation number for
$x_e \ll 1$ in the redshift range of interest, as the time scales for these
processes scale as $x_e^2$ (section~\ref{sec:tscale}). 
If photons are injected such that $T_B \gg T_e$ in a  small range of frequencies
for $x_e \ll 1$, Eqs.~(\ref{eq:double_comp}) and~(\ref{eq:freefree}) show that  the dominant process is the absorption of these photons. A fraction of this
absorbed energy heats electrons (Eq.~(\ref{eq:elec_temp})) and another fraction is re-emitted as soft photons. The loss of energy owing to expansion is not important as  the time scale of free-free processes, which are the dominant processes for $z <10^4$, is shorter than or comparable to the  expansion rate before recombination. 

We next consider the evolution of the  electron temperature (Eq.~(\ref{eq:elec_temp})). To simplify Eq.~(\ref{eq:elec_temp}), let us write photon occupation
number as:
\begin{equation}
  \eta(t,x_e) = \eta_0(x_e) + \delta\eta(x_e,t)
  \label{eq:backpch}
\end{equation}
Here $\eta_0(x_e)$ is the unperturbed distribution function at $T_{\rm CMB} = T_e$. We note that this split is not a perturbation expansion because $\delta\eta$ could dominate the equilibrium distribution for a range of frequency. However, electron temperature is determined by integrated quantities such as  photon energy  density which are dominated by equilibrium distribution function. Therefore, while such
a split would be less appropriate for studying the evolution of photon occupation number, it is suitable for studying the evolution of electron temperature. 
Using Eqs.~(\ref{eq:teq}) and~(\ref{eq:temp_ffdc}) and dropping the logarithmic frequency dependence of $g(x_e)$, we get:
\begin{equation}
  {d T_e \over dt} =   - {2\dot a \over  a}T_e + {(T_{\rm eq} - T_e) \over t_{\rm e\gamma}} + {T_e \over t_{\rm ff}} \int \delta\eta x_e dx_e + {T_e\over t_{\rm dc}} \int \delta\eta x_e dx_e
  \label{eq:elec_temp1}
\end{equation}
In writing the free-free and double Compton terms in Eq.~(\ref{eq:elec_temp1}) we have assumed that the support of $\delta\eta$ is in small range of frequencies
for $x_e \ll 1$. We expect the fractional  change in the temperature to be small
as compared to the unperturbed temperature, which allows us to replace temperature in $x_e$ with unperturbed temperature. Eq.~(\ref{eq:elec_temp1}) can then   be solved to give:
\begin{eqnarray}
  T_e(t) & \simeq   & T_e(t_i)\exp\left[-\int_{t_i}^t dt'\left ({1 \over t_{\rm e\gamma}} + {2\dot a \over a}+{1 \over t'_{\rm ff}}+{1 \over t'_{\rm dc}}\right )\right] + \exp\left[-\int_{t_i}^t dt'\left ({1 \over t_{\rm e\gamma}} + {2\dot a \over a}+{1 \over t'_{\rm ff}}+{1 \over t'_{\rm dc}}\right )\right] \nonumber\\
  &\times & \int_{t_i}^t dt' \exp\left[-\int_{t_i}^{t'} dt''\left ({1 \over t_{\rm e\gamma}} + {2\dot a \over a}+{1 \over t'_{\rm ff}}+{1 \over t'_{\rm dc}}\right )\right] {T_{\rm eq} \over t_{\rm e\gamma}}
  \label{eq:elec_sol}
\end{eqnarray}
Here we have redefined:
\begin{eqnarray}
  {1\over t'_{\rm ff}} & = & {1\over t_{\rm ff}} \int \delta\eta x_e dx_e \\
  {1\over t'_{\rm dc}} & = & {1\over t_{\rm dc}} \int \delta\eta x_e dx_e
\end{eqnarray}
For all the cases we consider, the integral over $\delta\eta$  is less than $10^{-6}$.  This,  in addition with  a comparison between different time scales, (Eqs.~(\ref{eq:invcomptemp}), (\ref{eq:fftemp}), (\ref{eq:dctemp}), and~(\ref{eq:exprate})), show that the evolution of the matter temperature is predominantly governed by the inverse  Compton scattering time scale. First, this means all terms except the inverse Compton term in Eq.~(\ref{eq:elec_sol}) can be dropped. As the inverse Compton
scattering rate far exceeds the expansion rate, the first term on the RHS of the equation can also be dropped. By making a change of variables in the second term, we get $T_e \simeq T_{\rm eq}$ (see e.g. \cite{1990PhRvD..41..354B}
for a similar physical setting). $T_{\rm eq}$ can be computed by solving the evolution of perturbed photon occupation number using Kompaneets equation (Eqs.~(\ref{eq:teq}) and~(\ref{eq:kompanee})). In all the cases we consider, $T_e-T_{e0} \ll T_{e0}$, where $T_{e0}$ is the unperturbed temperature.

Using the expansion given by Eq.~(\ref{eq:backpch}), we get:
\begin{equation}
  T_{\rm eq} = {T_{e0}\int_0^{\infty} dx_e x_e^4 \left (\eta_0 (\eta_0+1) + \delta\eta (\eta_0+1) + \delta\eta^2 \right) \over 4\int_0^{\infty} dx_e  x_e^3 \left(\eta_0 + \delta\eta \right)}
  \label{eq:pert_temp}
\end{equation}

\section{Numerical Solution of Kompaneets equation} \label{sec:numsol}
As discussed in the last section, our aim is to study the evolution of
the photon occupation number if the energy density
of injected  photons dominates the equilibrium distribution function, $T_{\rm B} \gg T_e$, for a small range of frequencies with $x_e \ll 1$. We discussed possible
analytical solutions in such a case taking into account Compton and inverse Compton scattering. We also showed that electron temperature relaxes to an effective radiation temperature given by Eq.~(\ref{eq:pert_temp}), which can be defined for arbitrary radiation fields, on time scales much shorter than all the  other time scales  (section~\ref{sec:tscale}). This allows us to split the  problem of simultaneous determination of photon occupation number and electron temperature into one in which slower processes determine the photon occupation number while the electron temperature varies according to $T_e = T_{\rm eq}$.

In this section, we seek numerical solutions to Eq.~(\ref{eq:kompanee}).  It  is a one-dimensional
PDE in variables  $x_{e}$ and $y$ coupled to the evolution of electron
temperature (Eq.~(\ref{eq:elec_temp})) \footnote{We note that the use of variable
  $x_e$ could be misleading when these two equations are coupled as electron temperature enters the definition of $x_e$. We use the variable $x_e$ with unperturbed electron temperature in this case and redefine $x_e$ as $x_e = h\nu/(kT_{e0})$ and use the electron temperature explicitly in  Eq.~(\ref{eq:kompanee}). However as $T_e-T_{e0} \ll T_{e0}$, our solutions are insensitive to the use of either of the two variables}.
 We used MATLAB to solve Eq.~(\ref{eq:kompanee}), making
use of the inbuilt PDE solver (pdepe). In order to be able to use the inbuilt solver to
solve this equation, we  convert it into the following form:
\begin{equation}
c\left(x,t,u,\frac{\partial u}{\partial x}\right)\frac{\partial u}{\partial t}=x^{-m}\frac{\partial}{\partial x}\left(x^{m}f\left(x,t,u,\frac{\partial u}{\partial x}\right)\right)+s\left(x,t,u,\frac{\partial u}{\partial x}\right)\label{eq:13}
\end{equation}
The variable $\eta x_e^3$ is used for numerical stability.

We assume two initial profiles for solving the Kompaneets equation: Gaussian and the profile corresponding to a decay product (to be discussed in the next section). The initial  photon occupation number is given by $\eta(x_e,z_i) = \delta\eta(x_e,t_i) + \eta_0(x_e)$, where $\eta_0(x_e)$ is the equilibrium distribution   and $\delta\eta$ is  the profile of the injected energy.

We first consider the Gaussian profile,  which is given by:
\begin{equation}
  x_e^3 \delta\eta(x_e,z_i)= {A \over \sigma\sqrt{2\pi}}\exp\left[-(x_e - \mu)^2/(2\sigma^2)\right]
  \label{eq:gauprof}
\end{equation}
Here $z_i$ is the redshift at which the energy is injected. $A$ is an overall
normalization that determines the total amount of injected energy. In addition to $t_i$ and $A$, the mean and the standard deviation of the Gaussian, $\mu$ and $\sigma$ give the other two free parameters for our study.

We fix $\mu \simeq 2\times 10^{-3}$  as required  by the  EDGES detection. For $z_i \gg 10^5$, $y \gg 1$, which means Compton scattering   time scales are short enough  to thermalize the excess of photons. This motivate us to   consider  $z_i < 10^5$ for our study.  The range of $A$ is  fixed by the EDGES observation. For
$x_e \ll 1$, $\eta \simeq 1/x_e$ for a Planckian.  $A$ is chosen such that the brightness temperature of  injected photons, $T_B \gg T_e$, as the EDGES detection   requires  the excess of
photons to be at least  a factor of 2.5 more than the  unperturbed CMB spectrum at $x_e \simeq 2\times 10^{-3}$.

\begin{figure}
        \centering
        \begin{minipage}{0.49\textwidth}
                \centering
                \includegraphics[width=1.0\linewidth]{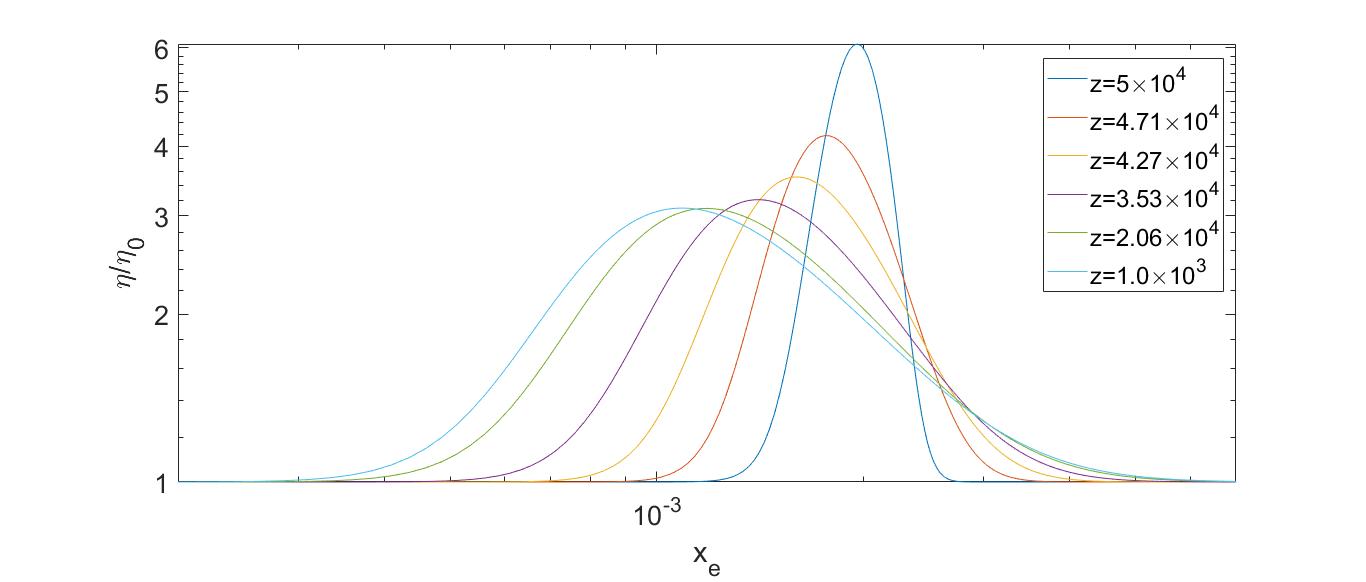}
        \end{minipage}\hfill
        \centering
        \begin{minipage}{0.49\textwidth}
                \centering
                \includegraphics[width=1.0\linewidth]{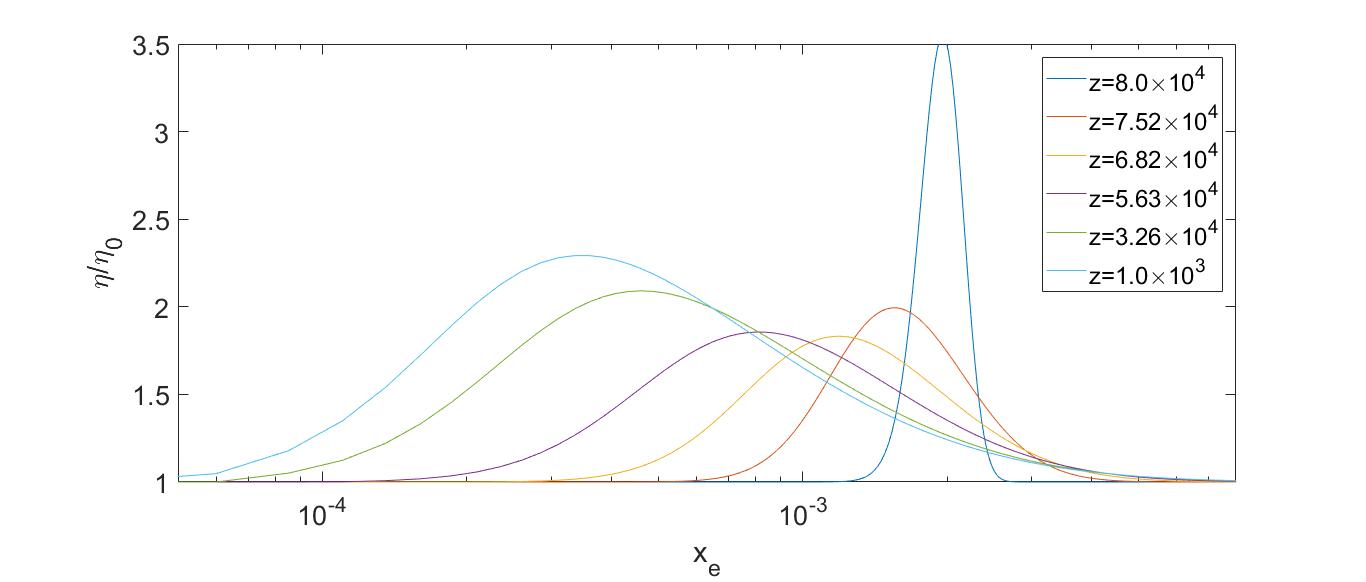}
        \end{minipage}\hfill
        \caption{The figures display the solution of Eq.~(\ref{eq:comp_invcomp}) for models in which a Gaussian profile is injected with mean $x_e = 2 \times 10^{-3}$ at two different redshifts.  The y-axis displays the ratio of the perturbed  and  the equilibrium CMB   occupation number  while the x-axis is in units $x_e \equiv h\nu/(kT_{e0})$. 
}
\label{fig:comp_scat}
\end{figure}

\begin{figure}
        \centering
        \begin{minipage}{0.49\textwidth}
                \centering
                \includegraphics[width=1.0\linewidth]{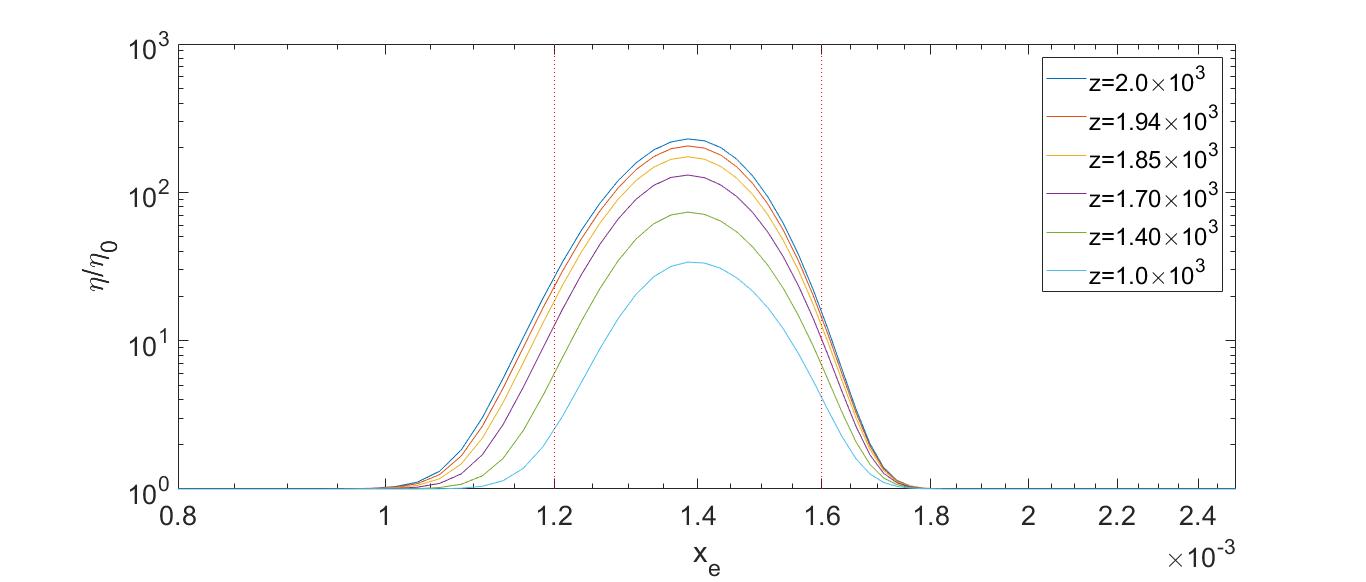}
        \end{minipage}\hfill
        \centering
        \begin{minipage}{0.49\textwidth}
                \centering
                \includegraphics[width=1.0\linewidth]{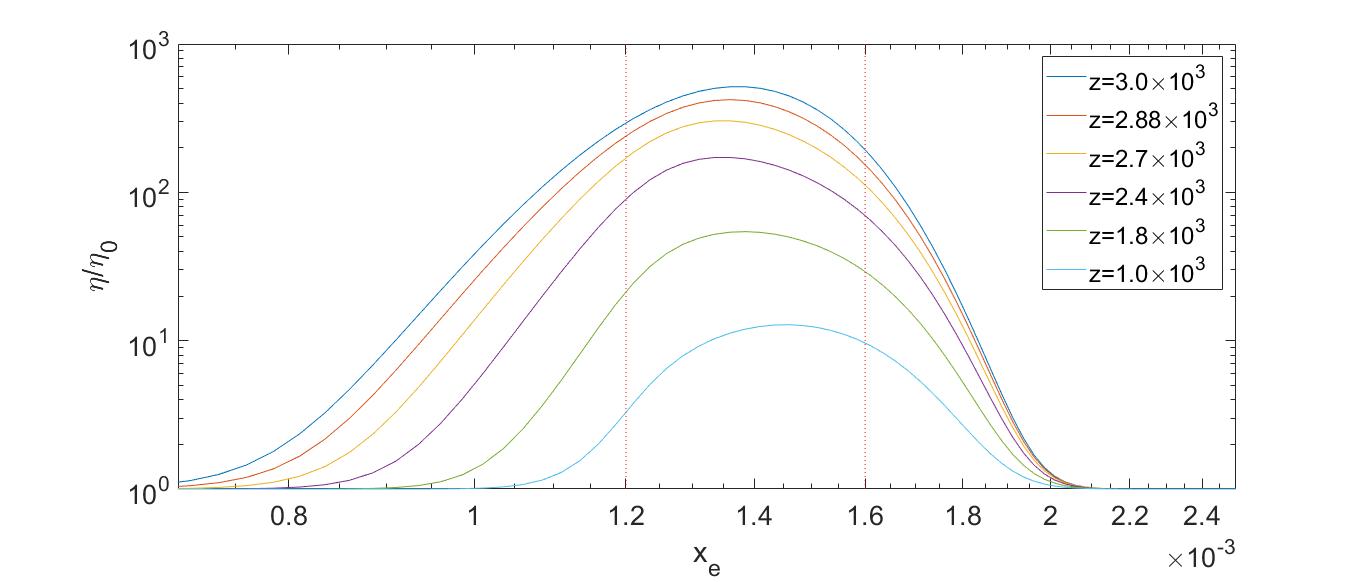}
        \end{minipage}\hfill
        \centering
        \begin{minipage}{0.49\textwidth}
                \centering
                \includegraphics[width=1.0\linewidth]{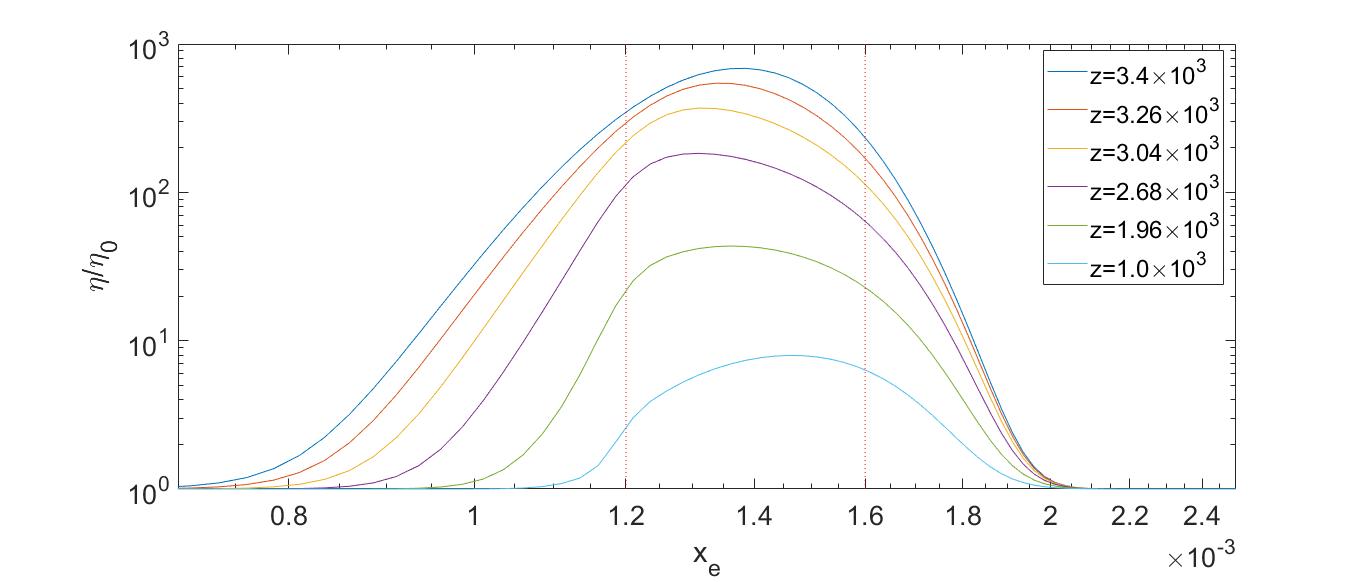}
        \end{minipage}\hfill
        \centering
        \begin{minipage}{0.49\textwidth}
                \centering
                \includegraphics[width=1.0\linewidth]{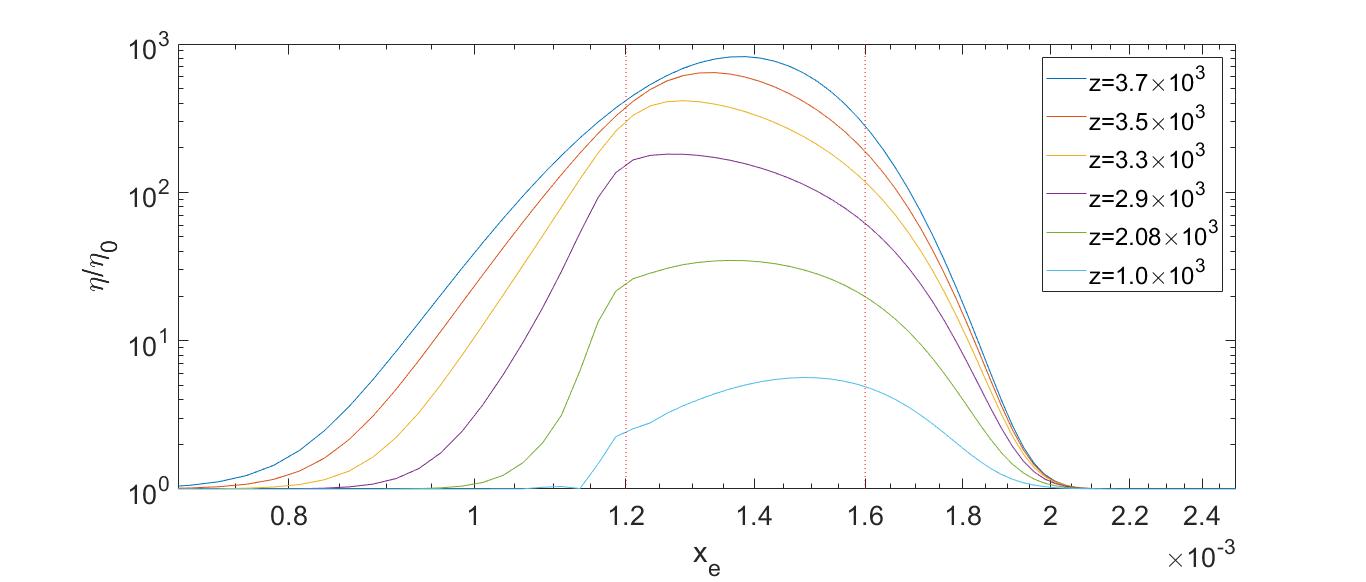}
        \end{minipage}\hfill
        \caption{For Gaussian input profiles of different amplitude, width, and  starting redshift, the panels show the evolution of the profile, including all
          the relevant physical processes (Eq.~(\ref{eq:kompanee})). The axes display the same quantities as in Figure~\ref{fig:comp_scat}.   The vertical lines correspond to $x_e = \{1.2, 1.6\}\times 10^{-3}$. To explain the  EDGES detection, the ratio displayed on y-axis  should exceed 2.5 in this range of $x_e$. 
}
\label{fig:allproc}
\end{figure}

To gauge the importance  of different physical
processes, we  show results for two different  cases: with only Compton/inverse Compton scattering and including  all the physical effects (Compton, double Compton and free-free processes).
In Figure~\ref{fig:comp_scat},  we show numerical solutions to Eq.~(\ref{eq:kompanee})
for  only Compton processes. The evolution of the profile is displayed for
a range of redshifts with $z< 8\times 10^4$. The profile stops evolving
at $z\simeq 1000$  as the universe recombines which causes   all the relevant time scales to become much larger than the expansion rate. 

Figure~\ref{fig:comp_scat} shows that the numerical
results agree with the analytic solutions (Eq.~(\ref{eq:20})): the Gaussian profile moves to lower energies with the steepening of the profile
as the dominant process  corresponds to stimulated emission  ($\eta^2$ term in
Eq.~(\ref{eq:comp_invcomp})). As the profile evolves and/or the injection energy
is pumped at an earlier era, the $\partial\eta/\partial x$ (inverse Compton scattering) starts playing a  role,  leading to up-scattering of the low energy photons. As  $y <1$ and  the excess photons  upscatter to frequencies  $x_f \simeq  x_i \exp(y)$, the inverse Compton scattering creates a tail above the frequencies of injection in the profile, as is seen in Figure~\ref{fig:comp_scat}. 
The net
impact of both these effects is to broaden the Gaussian, keeping invariant the
number density of excess photons $\propto \eta x_e^3$  while losing energy ($\propto \eta x_e^4$)  through Compton scattering to electrons. 

In Figure~\ref{fig:allproc} we include all the processes. For a photon excess,  double Compton/free-free processes mainly act as sink of photons. This is clearly seen when Figures~\ref{fig:allproc} and~\ref{fig:comp_scat} are compared. With only Compton processes  the initial distribution moves to smaller frequencies
with a small tail at frequencies larger than injected frequencies owing to
inverse Compton scattering. However, most of these photons are absorbed by free-free absorption which dominates at $z<10^4$. It should also be noted from  Figure~\ref{fig:allproc} that free-free absorption continues to deplete photons until $z\simeq 1000$ as its time scale
at $x \simeq 10^{-3}$ remains comparable to the expansion time scale until recombination. This also means that the EDGES result, which requires a minimum  residual excess of 2.5 at $x_e \simeq 2\times 10^{-3}$,  can generically be
achieved in the pre-recombination era only if the amount of  initial injection far exceeds the minimum required   residual excess, as is clear from  Figure~\ref{fig:allproc}. 

Figure~\ref{fig:allproc}  shows that to explain the EDGES detection
the energy injection redshift  $z_i < 4\times 10^3$.
For energy injection at higher redshifts, free-free processes    equilibrates  the surplus photons. It could be examined whether a larger amount of  injection could potentially leave a residue. We find
 this conclusion to be  insensitive to  the amount of injected energy  because this also increases the rate of absorption  (Eq.~(\ref{eq:freefree}) and~(\ref{eq:double_comp})). 

\subsection{Decaying particle}

In this section, we consider particle decay as the mechanism of pumping photons
in the CMB  at $x_e \ll 1$. In this model, a non-relativistic particle of
mass $m_{\rm d}$ decays into a neutral, massless particle and a photon. To get the photon injection in the desired frequency range, we need to assume that
  the mass of the particle is such that $m_{\rm d}c^2 \ll k T_e$. If this particle was coupled to the thermal bath of the universe in the early universe, it would be highly relativistic. Therefore, our implicit assumption is that
  this particle was always decoupled from the thermal bath  (e.g. an axion). For our work we require $m_{\rm d} \gtrsim 3 \times 10^{-3} \, \rm eV$. 
The occupation number  of  decay photons could be written as, in radiation dominated era (for details see \cite{1990PhRvD..41..354B}):
\begin{equation}
  \eta_{\rm decay}(\nu,t)=\frac{h^2 B n_{\rm d}(\tau)}{2\pi (m_{\rm d}/2)^{2}\nu c}\left(\frac{\tau}{t}\right)^{1/2}\exp\left({-\frac{t}{\tau}\left(\frac{2h\nu}{m_{\rm d} c^2}\right)^{2}}\right)\Theta\left(\frac{m_{\rm d}c^2}{2}-h\nu\right)
  \label{eq:decayspectrum}
\end{equation}
Here   $B$ is the branching ratio of photon decay. The initial conditions are set in the radiation dominated era; we make the switch to the relevant expression in the matter dominated era at matter radiation equality (for details see \cite{1990PhRvD..41..354B}). We can compute the ratio of the photon occupation number of  decay photons and CMB photons for $x_e \ll 1$. Using $n_{\rm d}(0) m_{\rm d}/\rho_c(0) = \Omega_{\rm d}$  and,   for CMB for
$x_e \ll 1$, $\eta_0 = 1/x_e$, we get, in the radiation dominated era:
\begin{equation}
  {\eta_{\rm decay}(\nu,t) \over \eta_0}=\frac{2 B h^3 \rho_c(0)\Omega_{\rm d} (1+z_d)^3}{\pi m_{\rm d}^{3}c (kT_e)}\left(\frac{\tau}{t}\right)^{1/2}\exp\left({-\frac{t}{\tau}\left(\frac{2h\nu}{m_{\rm d} c^2}\right)^{2}}\right)\Theta\left(\frac{m_{\rm d}c^2}{2}-h\nu\right)
  \label{eq:decaypro}
\end{equation}
Here $z_d$ is the redshift corresponding to $t = \tau$. It should be noted that
$\eta_{\rm decay}(\nu,t)/\eta_0$  is independent of both redshift  and frequency for $t \ll \tau$. 

It is easy to verify
that our requirement,  $\eta_{\rm decay}(\nu,t)/\eta \gg 1$, can readily be
achieved for a wide range of $m_{\rm d}$ and
$\tau$ for $\Omega_{\rm d} \ll 1$.

\begin{figure}
        \centering
        \begin{minipage}{0.49\textwidth}
                \centering
                \includegraphics[width=1.0\linewidth]{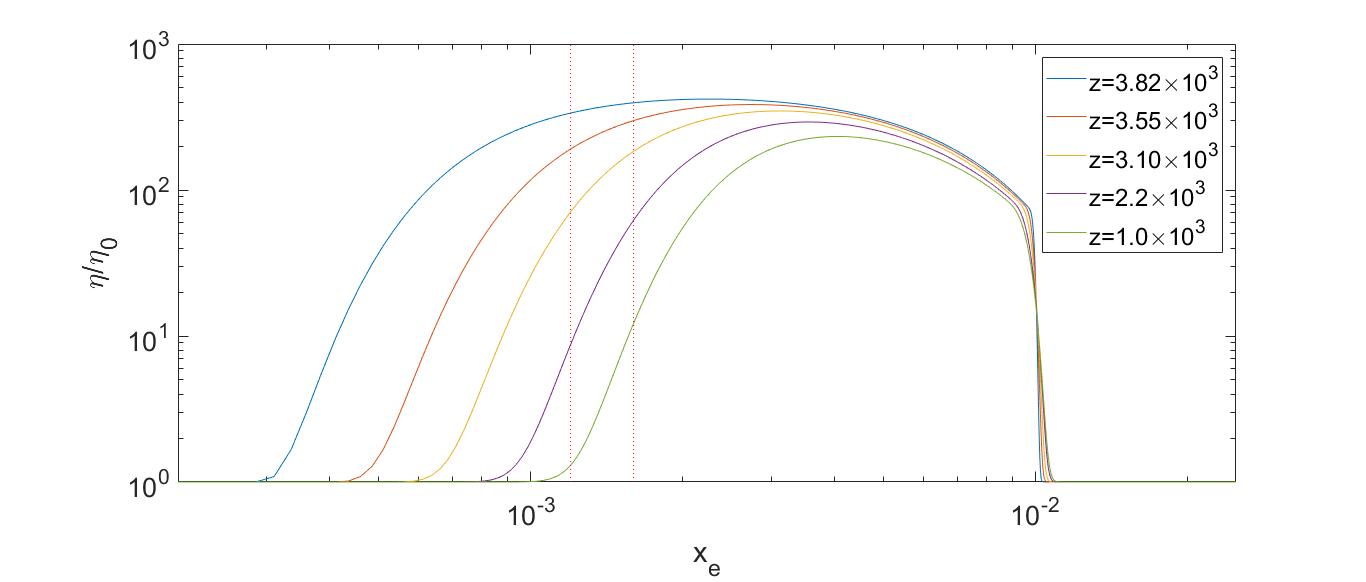}
        \end{minipage}\hfill
        \centering
        \begin{minipage}{0.49\textwidth}
                \centering
                \includegraphics[width=1.0\linewidth]{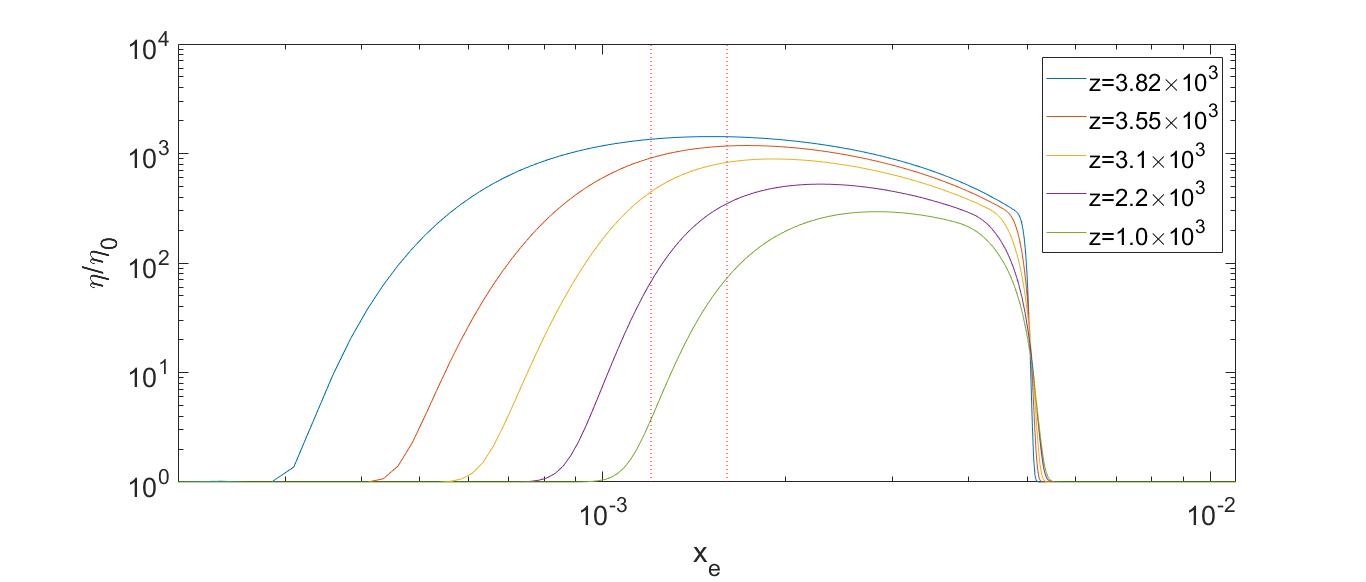}
        \end{minipage}\hfill
        \centering
        \begin{minipage}{0.49\textwidth}
                \centering
                \includegraphics[width=1.0\linewidth]{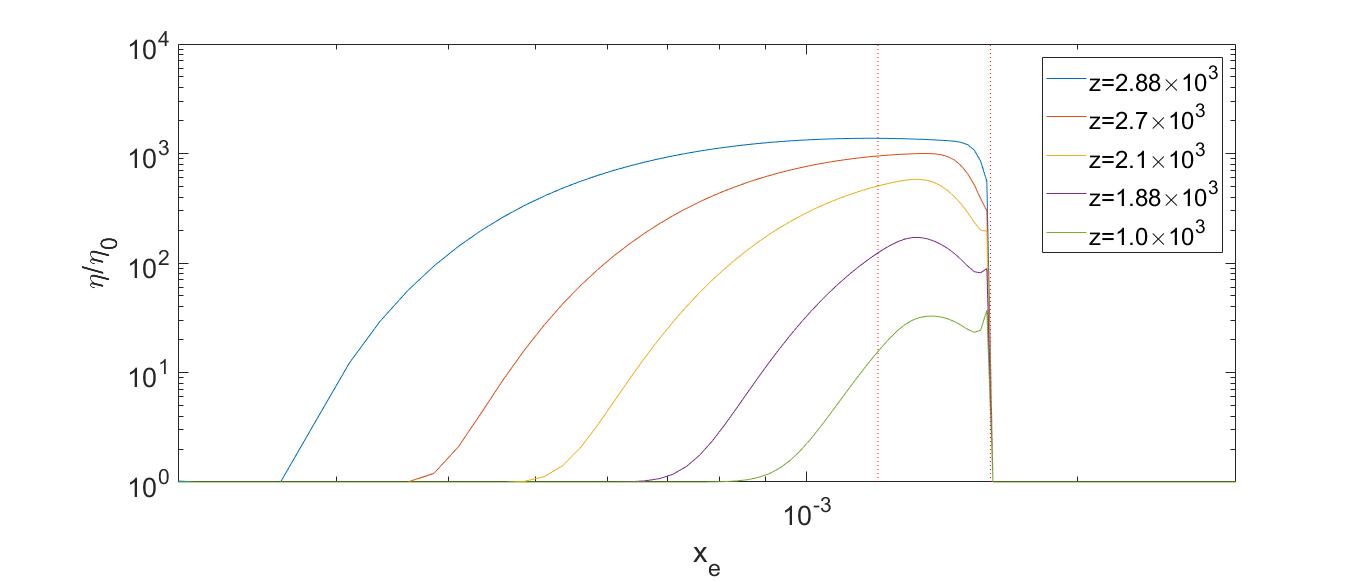}
        \end{minipage}\hfill
        \centering
        \begin{minipage}{0.49\textwidth}
                \centering
                \includegraphics[width=1.0\linewidth]{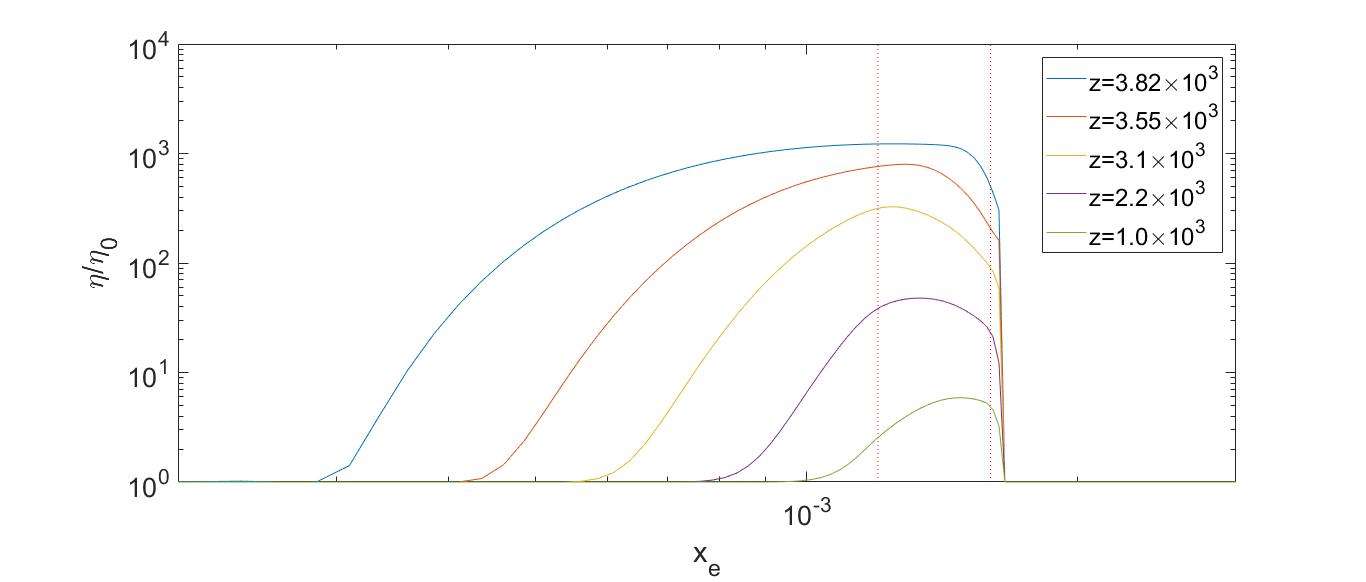}
        \end{minipage}\hfill
\caption{The evolution of the photon occupation number  is shown for decaying particle model (Eq.~(\ref{eq:decaypro})) for different values of decay redshift, $z_d$, and particle mass, $m_d$. The panels (clockwise from the top left) correspond to $z_d = \{5.6\times 10^3, 5.6\times 10^3, 2\times 10^3, 3\times 10^3\}$ and $m_d = \{2\times 10^{-2}, 9\times 10^{-3}, 3.1\times 10^{-3}, 3\times 10^{-3}\} \, \rm eV$, respectively. The axes display the same quantities as in Figure~\ref{fig:allproc}.
}
\label{fig:decayp}
\end{figure}

In Figure~\ref{fig:decayp}, the evolution of an input energy profile given
by Eq.~(\ref{eq:decaypro}) is shown for different decay times $\tau$ and particle mass $m_d$. 

To understand the figure and its implication, the following two facts must
be noted. First, all photons have the frequency, $\nu = m_d c^2/h$, at the time of production. 
These photons subsequently free stream in the expanding universe, losing their energy owing to redshift. Therefore, at any fixed time,  the photons with frequency $\nu \ll m_d c^2/(2h)$ are old photons that are absorbed with a greater probability because the rate of absorption is greater at higher redshifts. Second, the photon production is a continuous process.  For
$t \ll \tau$, the number of decay photons, $\simeq n_d t/\tau \ll n_d$. A majority of these photons are absorbed and thermalized. While these photons do not
play a role in explaining the residual, they are responsible for spectral distortion of the CMB which will be discussed in the next section. The photons that contribute to the observed excess are produced at $t \simeq \tau$. Therefore, to explain the residual photons at $z \simeq 1000$, we only need to model the late time evolution
of the decay photons and  it suffices to evolve photons close to the energy $m_dc^2/2$
at times close to $\tau$. We choose $z_i = z_d$ as our initial condition. We note that our results are insensitive to the choice of $z_i$ so long as $z_i \gtrsim  z_d$  and we verify this
by running our code with different $z_i$ up to  $z_i/z_d = 10$.

\subsection{Explaining EDGES data}
In Section~\ref{sec:edgesre}, we discussed the EDGES result  and different
ways to interpret  it. One possible  explanation is to increase $T_{\rm CMB}$ in Eq.~(\ref{overallnorm}) by at least  a factor of 2.5 in the observed frequency range 70--90~MHz, which corresponds to  $x_e =\{1.2,1.6\}\times 10^{-3}$. The range is marked in Figures~\ref{fig:allproc} and~\ref{fig:decayp}. 

In the Gaussian model, the requisite ratio of the residual excess $T_B/T_e$  can be achieved by varying $z_{i}$  and $A$.  As noted above,  if the  redshift of injection $z_i > 5 \times 10^3$,  the free free absorption causes the injected feature  to be thermalized. For $z_{i} <  4\times10^{3}$ we can  get the desired ratio for the input Gaussian profile (Figure~\ref{fig:allproc}). Figure~\ref{fig:allproc}  further shows that $T_B/T_e \gtrsim 200$ for acceptable models. 

For the decaying particle model, we find that $m_d \simeq 3 \times 10^{-3} \, \rm eV$ and  $z_d < 4 \times 10^3$ to  explain the 
EDGES data (Figure~\ref{fig:decayp}). We note that the EDGES detection constrains both $z_d$ and $m_d$.

All the cases shown in Figure~\ref{fig:allproc} and three of the four
cases shown in Figure~\ref{fig:decayp} can explain the EDGES data. 

\section{CMB spectral distortion} \label{sec:specdist}

In this paper, we invoke spectral distortion of CMB at $x_e \simeq 10^{-3}$ to explain the EDGES result. In addition to these frequencies, physical processes in the pre-recombination era also impact other parts of the CMB spectrum, which we term  'global' and 'local' spectral distortion. 

The energy injected into the
plasma heats electrons which, through inverse Compton scattering, transfer a
part of their energy to the CMB\footnote{Our main focus is  $y$-distortion which is caused if the energy is injected at $z \lesssim 10^5$. In this case, the rate of Compton processes is smaller   than the expansion rate (Eq.~(\ref{eq:invcompdist})), which prevents  relaxation to an  equilibrium distribution (e.g. \cite{2014PTEP.2014fB107T,1990PhRvD..41..354B} and references therein)}. This results in 'global' spectral distortion
as the energy can be transferred to CMB photons of any frequency. The other 'local' spectral distortion is caused by  the  upscattering/reemission  of injected
photons  close to the  frequency of injected photons.

{\em global spectral distortion}: The global CMB distortion is a two stage process: transfer of a fraction of injected photon energy (at $x_e \simeq 10^{-3}$)  to electrons  followed by a transfer of a fraction of this  energy to CMB at other frequencies  (e.g. \cite{1990PhRvD..41..354B} for details).
The dominant physical process for the former is free-free absorption and
inverse Compton scattering for the latter. Eqs.~(\ref{eq:invcomptemp}) and~(\ref{eq:freefree_eabs}) show that the time scales of these processes are  shorter than
the expansion time  in the pre-recombination era. This means that the loss of energy due to the expansion of the universe during the energy exchange is insignificant.
As $T_e \simeq  T_{\rm eq}$, the thermal energy density of baryons $\simeq n_ekT$
is negligible as compared to the CMB energy density, which implies a negligible
amount of injected energy is retained by particles.

Therefore, it is reasonable to assume that, apart from the residual, unthermalized energy, all the injected energy is transferred to CMB. Figures~\ref{fig:allproc}
and~\ref{fig:decayp} show that the residual is generally a small fraction of
the total injected energy. Using Eq.~(\ref{eq:eneinj}) we can estimate the
the amount of energy  absorbed by  the CMB. We require $\eta/\eta_0 \simeq 2.5$, or $T_{\rm B}/T \simeq 2.5$ in Eq.~(\ref{eq:eneinj}), to explain the EDGES result. However,  Figure~\ref{fig:allproc} shows
that   a much larger amount of energy, $\eta/\eta_0 \simeq 200\hbox{--}1000$, has to be injected  to achieve the requisite residual.  This yields
$\Delta \rho_{\rm \scriptscriptstyle CMB}/\rho_{\rm \scriptscriptstyle CMB0} \simeq 10^{-7} \hbox{--}10^{-6}$, and  it is a generic prediction of the model.

The case of decaying particles
is more complicated as the injection is a continuous process. To capture the time dependence of the energy exchange, we assume that all the decay photons are absorbed at the redshift of production. This is a reasonable assumption
for photons close to $x_e \simeq 10^{-3}$  as seen in  Figure~\ref{fig:decayp}.  The energy density of the decay photons
is  $\rho_{\rm d\gamma}(z) \simeq m_{\rm d} c^2 n_d(z) (1-\exp(-t/\tau))/2$; $n_{\rm d}(z) = n(0)(1+z)^3 \exp(-t/\tau)$ is the number density of decay particles. At $t\ll \tau$,  $\rho_{\rm d\gamma}(z)$ scales as $(1+z)$ and, even though the transfer of energy from decay photons to CMB is more efficient, the ratio  of decay photons and  CMB energy density  scales $(1+z)^{-3}$. This means the fractional change in CMB energy is smaller at higher redshifts. The main
contribution to CMB distortion arises from $z \lesssim z_d$,  which allows us to compute the spectral distortion as in the previous case. The EDGES detection
requires $z_d \lesssim 4 \times 10^3$ and $\eta_{\rm decay}/\eta_0 \simeq 100\hbox{--}1000$, which corresponds to $\Omega_{\rm d} \simeq  10^{-6}\hbox{--}10^{-5}$ for $B=1$ (Eq.~(\ref{eq:decaypro}) and Figure~\ref{fig:decayp}). This yields fractional change in CMB energy density:  $\Delta\rho_{\rm \scriptscriptstyle CMB}/\rho_{\rm \scriptscriptstyle CMB0}\simeq 10^{-6}\hbox{--}10^{-5}$.

The COBE-FIRAS data  yields the best upper limit on the CMB spectral distortion: $\Delta \rho_{\rm \scriptscriptstyle CMB}/\rho_{\rm \scriptscriptstyle CMB0} < 6 \times 10^{-5}$ for $\nu \gtrsim 60 \, \rm GHz$. Our predicted spectral distortion does not  violate the COBE-FIRAS bounds,  and its detection   is  within the capabilities  of
the upcoming satellite mission PIXIE (\cite{Kogut:2011xw}).

 {\em local spectral distortion}: The  'local' CMB 
distortion corresponds to the redistribution of injected photons higher up  the frequency ladder  close to the frequency of injection.  This is of particular interest because CMB spectrum has been
determined to high precision upto $\nu \simeq 1.5 \, \rm GHz$, which is only roughly a factor of 20 more than the  energy of injection \cite{2011ApJ...734....6S}.\footnote{We refer to frequencies  $\nu = \{3.2, 1.42\} \, \rm GHz$ of ARCADE data here at which the extragalactic residual is $2.787 \pm 0.01 \, \rm K$,  $3.18\pm 0.516 \, \rm K$, respectively. We could also consider the data point at $\nu = 408 \, \rm MHz$ at which the extragalactic residual is $10.8 \pm 3.5 \, \rm K$. } The energy injection in the early universe can be constrained by
ARCADE results \cite{pospelov2018room}. Figures~\ref{fig:decayp} and~\ref{fig:allproc} show
we do not expect significant emission at frequencies much larger than the injection frequencies. We can explain it by considering the relevant physical processes. 

Three physical processes can cause an increase in the frequency, through re-emission or scattering,  of injected photons: inverse Compton scattering, free-free emission, and  double Compton emission. As the time scale of free-free/double Compton processes  scales as $x_e^2$, they are generally inefficient at
re-emitting photons at frequencies 10 times larger than the energy of injection \footnote{Even if the time scale of emission  is shorter than expansion time scale, the absorption time scales are shorter than emission time scales if
  there is a photon excess (Eqs.~(\ref{eq:freefree}) and~(\ref{eq:double_comp})). The net effect of these processes is to create an equilibrium photon occupation number of temperature $T_e$  at frequencies smaller than the $x_e$ at which these time scales are shorter than expansion time scale.} So the dominant process that re-distributes energy to higher frequency is inverse Compton scattering, as our numerical results show. As the rate of inverse Compton scattering is smaller than the  free free absorption and the process  raises the  frequency as  $\exp(y)$, we
do not expect significant excess at frequencies separated from the frequency of injection so long as $y \le 1$. 

This means the 'local' spectral distortion could be negligible even if
a large amount of energy is injected. Alternatively, we could seek
to explain, or put constraints from, ARCADE results by adjusting the time of injection of energy and for the decaying particle model, $m_d$. One of the panels of
Figure~\ref{fig:decayp}  (top left) shows a model in which ARCADE result at 408~MHz is violated. However, this model is not consistent with the  EDGES results for the following reason: for causing substantial distortion at frequencies probed by ARCADE result, $m_d$ needs to be much larger than the value required to explain the EDGES result. This means photons close to $x_e \simeq 1.5 \times 10^{-3}$ are old photons, as explained above, and are absorbed more efficiently at higher redshifts.

\subsection{Impact on CMB anisotropies}
The injection of photons at $x_e \ll 1$  could leave detectable imprints on CMB anisotropies for the following reasons:
\begin{itemize}
\item[1.] The change in the matter/radiation content of the universe  alters the matter-radiation equality. 
\item[2.] The  photons and baryons are tightly coupled before the recombination commences, as the time scale of photon-electron scattering, $t_c = 1/(n_e \sigma_T c)$, is shorter than the expansion time scale  during this era. The  extra
  radiation energy, tightly coupled to baryons,   results in lower baryon loading $R\equiv 3\rho_b/(4\rho_\gamma)$, which is strongly constrained by CMB anisotropies (for details see e.g. \cite{2003moco.book.....D} and references therein).
\end{itemize}
The impact of these effects can be studied within the framework of a specific
model, e.g. decaying particle scenario we outlined above. In this model,
a non-relativistic particle of density parameter $\Omega_{\rm d}$ contributes to the cold dark matter of the universe before it  decays into
radiation on a time scale  $\tau$.  The  relative impact  of this model  on CMB anisotropies  can be discerned
from  the range of permissible values of $B$ and  $\Omega_{\rm d}$ from   Planck CMB data.

From Eq.~(\ref{eq:decaypro}) it follows that, for $B=1$, the ratio of decay photons to CMB occupation number 
is unity for   $\Omega_{\rm d} \simeq 10^{-8} \Omega_m$ (for $z_d = 3 \times 10^3$), where $\Omega_m =  0.315\pm 0.007$ is the best-fit matter density parameter  from Planck.   To explain the EDGES result, we require $\Omega_{\rm d} \simeq 10^{-5}$ which is   well within the precision with which $\Omega_m$ has been determined by Planck. However, for $B\ll1$, the case for which only a small fraction of particles decay into  photons,  $\Omega_{\rm d}$  needs to be  larger by a factor $1/B$  to explain the EDGES result. It should be noted that  the amplitude of CMB  spectral
distortion also  depend on the product, $B \Omega_{\rm d}$. However,  CMB anisotropies are  sensitive to $\Omega_{\rm d}$ in addition. Therefore, CMB anisotropies  provide a   complementary probe of the decaying particle model.

Planck data measures the  angular acoustic scale with   0.03\% precision \cite{Planck2018}.
The angular acoustic scale  is   $\propto 1/(1+R)^{1/2}$. 
From Eq.~(\ref{eq:eneinj}) it follows that the additional energy $\delta\rho_\gamma \simeq 10^{-9} \rho_{\rm \scriptscriptstyle  CMB}$ for $T_B = T_e$. For explaining the EDGES result
we require $T_B/T_e \lesssim 1000$ which causes a fractional  change $\Delta R/R \simeq 10^{-6}$,  which is well within   Planck constraints. Similar conclusions can be reached for the decaying particle model.

Therefore, the scenario studied in this paper is consistent with Planck data.
However, decaying particle  models with $B \ll 1$  might yield  observable features in CMB data and we hope to return to this study in the future.

\section{Summary and conclusions}
In this paper we study the possibility of explaining the recent EDGES
detection using energy injection  in CMB at  $x_e \ll 1$ during the  pre-recombination era. We study, both analytically and numerically,  the evolution of the CMB spectrum in the
presence of energy injection of  arbitrary amplitude and  
redshift of injection. All the important physical processes---Compton/inverse Compton  scattering, free-free absorption/emission, and double Compton absorption/emission---are considered in our study. We analyze two different models of
energy injection: (a) Gaussian profile with varying amplitude and injection redshifts, (b) decay of a non-relativistic particle parameterized in terms of its number density,  mass, and  decay redshift.

We show that if the energy injection in the relevant frequency range ($x_e \simeq 2\times 10^{-3}$) occurs after $z\simeq 4 \times 10^3$, the
energy density of the   residual, unthermalized photons in the CMB can  explain the EDGES detection. Figures~\ref{fig:decayp} and~\ref{fig:allproc} also show that, for a generic injection event, the injected energy needs to be more than two orders of magnitude larger than the minimum  residue  required to explain the  EDGES detection. 

We find that the  energy injection would also cause: (a) 'global' y-distortion
of CMB owing to  the heating of  electrons, (b) 'local' distortion of the CMB owing to the upscattering of injected photons, (c) additional  CMB anistropies because of the  increased matter and radiation energy density.

The 'global' y-parameter is currently constrained by COBE-FIRAS observation to
be $\lesssim  10^{-5}$.
The upcoming instrument PIXIE will improve this by up to four orders of magnitude. The 'local' distortion is constrained by ARCADE observations. Planck results have precisely determined   both the radiation and matter energy density in the universe.  We show that our proposed scenario satisfies all these constraints. 
Therefore, the model we consider is tightly constrained by current observations and
make meaningful predictions for the future observations. In particular, PIXIE might be able to detect such an energy injection in the pre-recombination era.

It might be possible to distinguish the energy injection in the pre-recombination era from other mechanisms that  have been invoked to explain the  EDGES detection  in the post-recombination
era, e.g. production of photons owing to radio sources or the milli-charged dark matter particle. Figures~\ref{fig:decayp} and~\ref{fig:allproc} show that
there are no residual photons for $x_e < 10^{-3}$. This is a generic feature of the  pre-recombination physical processes as the   free-free absorption time scale  for  these photons  is shorter than the expansion time scale  even close to the epoch of recombination.  However, the low-redshift radio background is unlikely to have such a cut-off 
and, therefore, it  will also impact the pre-reionization EoR absorption feature at $\nu \simeq 30 \, \rm MHz$. The  proposed  mission FARSIDE, which is  capable of detecting the pre-reionization HI signal, will  be able to 
 distinguish between these models. \footnote{
https://www.lpi.usra.edu/leag/white-papers-astronomy/FARSIDE\_190710\_Final.pdf}

 While the main aim of this paper is to explain the EDGES detection, we have presented a general formalism to study the evolution of energy injection in the pre-recombination era at $x_e \ll 1$. It allows for the injected energy to far exceed the  CMB energy for a small range of frequencies  if the total amount of injected energy remains small as compared to the total CMB energy. 

 \section{Appendix: Time scales of various radiative processes} \label{sec:tscale}

 In this Appendix, we list  different time scales relevant for our study.

{\it Compton/inverse Compton scattering}: For Compton/inverse Compton scattering, an important time scale is the
energy relaxation time scale between electrons and photons if  electron
is the target. This time scale determines the evolution of 
the  photon occupation number:
\begin{equation}
  t_{\gamma e} = \left({1\over n_e(z) c \sigma_{\rm T}} \right ) \left ({m_e c^2 \over kT_{\rm CMB}} \right ) = 4.8 \times 10^{13} \left ({10^4 \over 1+z} \right)^4 \, {\rm sec}
  \label{eq:invcompdist}
\end{equation}
 Similarly, we can define a relaxation time scale if photon
is the target. This time scale determines the evolution of matter temperature. 
\begin{equation}
  t_{e\gamma}=\frac{3m_{e}c}{4\sigma_{\rm T}\rho_{\rm CMB}}=7.3 \times 10^3 \left ({10^4 \over 1+z} \right)^4 \, {\rm sec} \label{eq:invcomptemp}
\end{equation}

{\it Free-free emission/absorption}: The relevant time scale for the
evolution of photon occupation number  is $x_e^3\eta/(g(x_e)K_0)$ (Eq.~(\ref{eq:freefree})) with \footnote{see e.g. \cite{1977NCimR...7..277D}; $K_0$ is more readily derived by starting with the emissivity of thermal free-free emission (e.g. \cite{1994ApJ...427..603R}) and diving by $c^3/(8\pi h\nu^3)$ to convert from
  emissivity to  the time derivative of  the photon occupation number; this yields $K_0/x_e^3$.}
\begin{equation}
  K_0^{-1}(z) = \left ({ 8\pi^{1/2} e^6h^2n_e^2(z) \over (54)^{1/2} m_e^{3/2} (kT_{\rm CMB})^{7/2}} \right)^{-1} = 5.3 \times 10^{16} \left ({10^4 \over 1+z} \right)^{5/2} \, {\rm sec}
  \label{eq:freefree_eabs}
\end{equation}
For determining the electron temperature, the  time scale of interest  is (Eq.~(\ref{eq:temp_ffdc})):
\begin{equation}
  t_{\rm ff} = \left ( {8\pi (kT_{\rm CMB})^4 K_0 \over c^3 h^3 n_e(z) k T_{\rm CMB}} \right )^{-1} = 7 \times 10^7 \left ({10^4 \over 1+z} \right)^{5/2} \, {\rm sec}
  \label{eq:fftemp}
\end{equation}

{\it Double Compton emission/absorption}: The relevant time scale for the
evolution of the photon occupation number  is $x_e^3\eta/C(t)$ (Eq.~(\ref{eq:double_comp})) with
\begin{equation}
  C^{-1}(z)  = 3.2 \times 10^{19} \left ({10^4 \over 1+z} \right)^{5} \, {\rm sec}
\end{equation}
For determining the electron temperature, the  time scale is (Eq.~(\ref{eq:temp_ffdc})):
\begin{equation}
  t_{\rm dc} = \left ( {8\pi (kT_{\rm CMB})^4 C(t) \over c^3 h^3 n_e(z) k T_{\rm CMB}} \right )^{-1} = 1.1 \times 10^{12} \left ({10^4 \over 1+z} \right)^{5} \, {\rm sec}
  \label{eq:dctemp}
\end{equation}

{\it Expansion time scale}: It is given by the inverse of the expansion rate, $H^{-1}(z)$:
\begin{equation}
  t_{\rm exp} = {H^{-1}_0 \over \left (\Omega_m (1+z)^3 +\Omega_r (1+z)^4 \right)^{1/2}} \simeq 4.6 \times 10^{11} \left ({10^4 \over 1+z} \right)^{2} \, {\rm sec}
  \label{eq:exprate}
\end{equation}
The numerical value quoted in Eq.~(\ref{eq:exprate}) correspond to radiation-dominated era, using $\Omega_r = 4.3 \times 10^{-5}$. 

Finally, we often  use  the Compton $y$ parameter as the dimensionless time
variable. It is  given by:
\begin{equation}
y(t)=\intop_{t_i}^{t}dt' a_{c}\label{eq:11}
\end{equation}
where $t_i$ is the time of energy injection and $a_{c}^{-1}$ is the characteristic time scale
on which the radiation spectrum reaches a state of quasi-equilibrium
under Compton/inverse Compton  scattering alone. A useful expression for numerical  computation is: 
\begin{equation}
{dy \over dz}= -8.66\times10^{-11}\frac{(1+z)^{3/2}}{\sqrt{5725+(1+z)}}
\end{equation}

The  time scales discussed above could change substantially  when the  departure
from the equilibrium state is large. In this paper we study  a scenario in which the injected number of photons  far exceeds the number  of  CMB photons in
equilibrium for  $x_e \ll 1$.  This means the time scales for the evolution of photon occupation number for $x_e \ll 1$ could  depend on the
number of injected photons. If we assume the brightness temperature in
a small frequency range is $T_B$ such that $T_B \gg T_e$, it follows from  Eqs.~(\ref{eq:double_comp}) and~(\ref{eq:freefree}) that the relevant times scales for the evolution of
the photon occupation number   are  $\simeq x_e^3\eta_0 T_e/(C(t)T_B)$ and $\simeq x_e^3\eta_0T_e/(K_0T_B)$ for double Compton and free-free processes, respectively. Similarly, as  we see in section~\ref{sec:anaappro}, the
time scale at which photons lose energy to electron through Compton scattering  is inversely proportional to the amount of injected energy (Eq.~(\ref{eq:20})). However, the times scales of the evolution of electron temperature, determined by Eqs.~(\ref{eq:dctemp}) and~(\ref{eq:fftemp}), are representative as the total amount of energy injected is small as compared to the total CMB energy density.

\bibliography{bib}
\end{document}